\documentclass[a4paper,fleqn,usenatbib,useAMS]{mnras}
\usepackage{graphicx}   % Including figure files
\usepackage{color}
\usepackage{lscape}
\usepackage{txfonts}
\usepackage{amsfonts}

\title[Radiative efficiency at low-level accretion]
{The radiative efficiency of neutron stars at low-level accretion} 
\author[Erlin Qiao and B.F. Liu]{Erlin Qiao $^{1,2}$\thanks{E-mail:
qiaoel@nao.cas.cn} and B.F. Liu $^{1,2}$\\
$^{1}$Key Laboratory of Space Astronomy and Technology, National Astronomical Observatories, Chinese Academy of
Sciences, Beijing 100101, China \\
$^{2}$School of Astronomy and
Space Sciences, University of Chinese Academy of Sciences, 19A Yuquan Road, Beijing 100049, China\\} 

\date{Accepted XXX. Received YYY; in original form ZZZ}
\pubyear{2020}

\begin{document}
\label{firstpage}
\pagerange{\pageref{firstpage}--\pageref{lastpage}}
\maketitle
%--------------------------------------------------------------------------------------------
% Abstract of the paper
\begin{abstract}
When neutrons star low-mass X-ray binaries (NS-LMXBs) are in the low-level accretion regime 
(i.e., $L_{\rm X}\lesssim 10^{36}\ \rm erg\ s^{-1}$),  
the accretion flow in the inner region around the NS is expected to be existed in the form of the hot accretion 
flow, e.g., the advection-dominated accretion flow (ADAF) as that in black hole X-ray binaries. Following our 
previous studies in Qiao \& Liu 2020a and 2020b on the ADAF accretion around NSs, in this paper, we investigate 
the radiative efficiency of NSs with an ADAF accretion in detail, showing that the radiative efficiency of NSs 
with an ADAF accretion is much lower than that of $\epsilon \sim {\dot M GM\over R_{*}}/{\dot M c^2}\sim 0.2$ 
despite the existence of the hard surface. 
As a result, given a X-ray luminosity $L_{\rm X}$ (e.g., between 0.5 and 10 keV), $\dot M$ calculated by 
$\dot M=L_{\rm X}{R_{*}\over {GM}}$ is lower than the real $\dot M$ calculated within the framework of the 
ADAF accretion. The real $\dot M$ can be more than two orders of magnitude higher than that of calculated by 
$\dot M=L_{\rm X}{R_{*}\over {GM}}$ with appropriate model parameters.
Finally, we discuss that if applicable, the model of ADAF accretion around a NS can be applied to 
explain the observed millisecond X-ray pulsation in some NS-LMXBs (such as PSR J1023+0038, XSS J12270-4859
and IGR J17379-3747) at a lower X-ray luminosity of a few times of $10^{33}\ \rm erg\ s^{-1}$,
since at this X-ray luminosity the calculated $\dot M$ with the model of ADAF accretion can be high enough to 
drive a fraction of the matter in the accretion flow to be channelled onto the surface of the NS
forming the X-ray pulsation.
\end{abstract}

% Select between one and six entries from the list of approved keywords.
% Don't make up new ones.
%\begin{keywords}
%keyword1 -- keyword2 -- keyword3
%\end{keywords}

\begin{keywords}
accretion, accretion discs
-- stars: neutron 
-- black hole physics
-- X-rays: binaries
\end{keywords}

%%%%%%%%%%%%%%%%%%%%%%%%%%%%%%%%%%%%%%%%%%%%%%%%%%%%%%%%%%%%%%%%%%%%%%%%%%
%%%%%%%%%%%%%%%%% BODY OF PAPER %%%%%%%%%%%%%%%%%%%%%%%%%%%%%%%%%%%%%%%%%%
\section{Introduction}
Currently, there are two types of accretion flow around compact objects 
[black holes (BH) and neutron stars (NS)], i.e., the geometrically thin, optically thick, cold accretion disc 
\citep[][]{Shakura1973}, and the geometrically thick, optically thin, hot accretion 
flow, such as the advection-dominated accretion flow (ADAF) \citep[][for review]{Yuan2014}.
The cold accretion disc with a higher mass accretion rate is widely used to explain the optical/UV emission 
in luminous active galactic nuclei (AGNs), and the X-ray emission of X-ray binaries at the high/soft 
state \citep[e.g.][]{Mitsuda1984,Makishima1986}. 
While the ADAF with a lower mass accretion rate is often used to explain the dominant emission in 
low-luminosity AGNs, as well as the low/hard and the quiescent state of X-ray 
binaries \citep[][for review]{Done2007}.

In general, the cold accretion disc is a kind of radiatively efficient accretion flow in both BH case
and NS case. In the approximation of the Newtonian mechanics, for a non-rotating BH, the radiative efficiency of 
the accretion disc is $\epsilon={1\over 2}{\dot M GM\over 3R_{\rm S}}/{\dot M c^2}\sim 0.1$ 
(with $G$ being the gravitational constant, $\dot M$ being the mass accretion rate in units 
of $\rm g\ s^{-1}$, $c$ being the speed of light, and $R_{\rm S}$ being the Schwarzschild radius 
with $R_{\rm S}=2GM/c^2\approx 2.95\times 10^{5}\ M/M_{\odot}\ \rm cm$), i.e., half of the gravitational 
energy will be released out in the form of the electromagnetic radiation in the accretion disc. 
%---------------------------
While for a NS, the radiative efficiency of the accretion is 
$\epsilon={\dot M GM\over R_{*}}/{\dot M c^2}\sim 0.2$ (taking $M=1.4M_{\odot}$, and $R_{*}=12.5\ \rm km$
\footnote{We take $R_{*}=12.5$ because we 
intend to keep the same efficiency of the gravitational energy release between a NS and a non-rotating BH. 
If the NS mass $m=1.4$ is taken, the corresponding NS radius is $R_{*}=12.5$ km.}), 
which is obtained as that half of the gravitational energy will be released out in the accretion disc and 
the other half of the gravitational energy will be released out in a thin boundary layer between the accretion 
disc and the surface of the NS in the form of the electromagnetic radiation \citep[][]{Gilfanov2014}.

In general, the ADAF solution is a kind of radiatively inefficient accretion flow in BH case.
In the approximation of the Newtonian mechanics, for a non-rotating BH, the radiative efficiency 
of the ADAF is $\epsilon<{1\over 2}{\dot M GM\over 3R_{\rm S}}/{\dot M c^2}\sim 0.1$
\citep[][for review]{Ichimaru1977,Rees1982,Narayan1994,Narayan1995b,Manmoto1997,Yuan2014}.
This is due to the optically thin nature of the ADAF solution, a fraction of the viscously dissipated 
energy in the ADAF is stored in the gas of the ADAF as the internal energy, and finally advected into the 
event horizon of the BH without radiation. 
%--------------------------
The fraction of the viscously dissipated energy stored in the gas of the ADAF as the internal energy is dependent 
on $\dot M$, and this fraction increases with decreasing $\dot M$, which means that the radiative efficiency 
of the ADAF around a BH decreases with decreasing $\dot M$. Specifically, when $\dot M$ is close the critical 
mass accretion rate $\dot M_{\rm crit}$ of the ADAF ($\dot M_{\rm crit} \sim \alpha^2\dot M_{\rm Edd}$, 
with $\alpha$ being the viscosity parameter, and $\dot M_{\rm Edd}=L_{\rm Edd}/{0.1c^2}
\approx 1.39\times 10^{18} M/M_{\odot}\ \rm g \ s^{-1}$ being the 
Eddington scaled mass accretion rate, where $L_{\rm Edd}$ is defined as 
$L_{\rm Edd}=1.26\times 10^{38} M/M_{\odot}\ \rm {erg\ s^{-1}}$), the value of $\epsilon$ is close to 0.1 as 
that of the cold accretion disc around a BH \citep[][]{Xie2012}. However, when $\dot M$ is significantly less 
than $\dot M_{\rm crit}$, the radiative efficiency $\epsilon$ decreases dramatically with decreasing $\dot M$, and 
the value of $\epsilon$ is much less than 0.1 \citep[][for discussions]{Xie2012}.

%----------
In this paper, we focus on the radiative efficiency of 
the ADAF solution around a NS (strictly speaking, it is the radiative efficiency of NSs with an ADAF accretion,
since a fraction of the ADAF energy released at the surface of the NS can finally radiate out to be observed). 
The dynamics of the ADAF around a weakly magnetized NS has been investigated by some authors 
previously \citep[][]{Medvedev2001,Medvedev2004,Narayan1995b}. In general, one of the most difficult problems 
for the study of the ADAF around a NS is how to treat the dynamics and radiation of the boundary layer between 
the surface of the NS and the ADAF. The physics of the boundary layer can significantly affect the global dynamics 
and radiation of the ADAF \citep[][]{Medvedev2001,Medvedev2004,DAngelo2015}.  
Recently, in a series of papers, i.e., \citet[][]{Qiao2018b}, \citet[][]{Qiao2020a}, 
and \citet[][]{Qiao2020b}, we study the dynamics and the radiation of the ADAF around a weakly magnetized NS 
in the framework of the self-similar solution of the ADAF by simplifying the physics of the boundary layer. 
Specifically, we introduce a parameter, $f_{\rm th}$, which describes the fraction of the ADAF energy 
released at the surface of the NS as thermal emission to be scattered in the ADAF. Under this assumption, i.e., 
considering the radiative feedback between the surface of the NS and the ADAF, we self-consistently calculate 
the structure and the corresponding emergent spectrum of the NS with an ADAF accretion. 
%------
The value of $f_{\rm th}$ can affect the radiative efficiency of NSs with an ADAF accretion.
Physically, the value of $f_{\rm th}$ is uncertain. However, it has been shown that the value of $f_{\rm th}$ 
can be constrained in a relatively narrow range by comparing with the observed X-ray spectra 
(typically between 0.5 and 10 keV) of neutron star low-mass X-ray binaries (NS-LMXBs), 
since the value of $f_{\rm th}$ can affect both the shape and the luminosity of the X-ray 
spectra \citep[][]{Qiao2020a,Qiao2020b}.
The results in \citet[][]{Qiao2020a,Qiao2020b} jointly suggest that the value of 
$f_{\rm th}$ is certainly less than 0.1, and a smaller value of $f_{\rm th}\sim 0.01$ is
more preferred.

In this paper, following the results of \citet[][]{Qiao2020a,Qiao2020b} for the constraints to the 
value of $f_{\rm th}$, we investigate the radiative efficiency of NSs with an ADAF accretion for taking two 
typical values of $f_{\rm th}$ as that of $f_{\rm th}=0.1$, and $f_{\rm th}=0.01$ respectively.
The radiative efficiency is defined as $\epsilon_{\rm bol}=L_{\rm bol}/{\dot M c^{2}}$ (with $L_{\rm bol}$ 
being the bolometric luminosity). Based on the emergent spectra of NSs with an ADAF accretion for the 
bolometric luminosity, we find that $\epsilon_{\rm bol}$ is nearly a constant with $\dot M$ for either 
taking $f_{\rm th}=0.1$ or taking $f_{\rm th}=0.01$.
The value of $\epsilon_{\rm bol}$ for $f_{\rm th}=0.1$ is roughly one order of magnitude lower than the 
previously expected value of $\epsilon \sim {\dot M GM\over R_{*}}/{\dot M c^2}\sim 0.2$, and 
$\epsilon_{\rm bol}$ for $f_{\rm th}=0.01$ is roughly two orders of magnitude lower than the value of 
$\epsilon \sim 0.2$. Then, we suggest that NSs with an ADAF accretion is radiatively inefficient 
despite the existence of the hard surface. 
%---------
Further, we investigate the radiative efficiency in some specific bands, e.g., 
$\epsilon_{\rm 0.5-10keV}$ and $\epsilon_{\rm 0.5-100keV}$ (defined as $\epsilon_{\rm 0.5-10keV}=L_{\rm 0.5-10keV}/{\dot M c^{2}}$
and $\epsilon_{\rm 0.5-100keV}=L_{\rm 0.5-100keV}/{\dot M c^{2}}$).
$\epsilon_{\rm 0.5-10keV}$ is nearly same with $\epsilon_{\rm 0.5-100keV}$ for a fixed $\dot M$ with 
$f_{\rm th}=0.1$ and $f_{\rm th}=0.01$ respectively.
%---------
$\epsilon_{\rm 0.5-10keV}$ (or $\epsilon_{\rm 0.5-100keV}$) is less than $\epsilon_{\rm bol}$, so is certainly less than  
$\epsilon \sim {\dot M GM\over R_{*}}/{\dot M c^2}\sim 0.2$. 
Meanwhile, $\epsilon_{\rm 0.5-10keV}$ (or $\epsilon_{\rm 0.5-100keV}$) decreases very quickly with decreasing $\dot M$.
As a result, for a NS-LMXB, if we intend to use the observed X-ray luminosity (e.g., between 0.5 and 10 keV) 
as the indicator for $\dot M$, $\dot M$ calculated with the formula of $\dot M=L_{\rm X}{R_{*}\over {GM}}$ is lower 
than that of calculated with our model of ADAF accretion around a NS. Obviously, given a X-ray luminosity, the 
difference between the $\dot M$ calculated with our model of ADAF accretion and the $\dot M$ calculated with the 
formula of $\dot M=L_{\rm X}{R_{*}\over {GM}}$ depends on $f_{\rm th}$, with the difference of the calculated 
$\dot M$ increases with decreasing $f_{\rm th}$.

Finally, in this paper, we argue that if applicable, the model of ADAF accretion around a NS can probably be used 
to explain the observed millisecond X-ray pulsation in some NS-LMXBs (such as PSR J1023+0038, XSS J12270-4859
and IGR J17379-3747) at a X-ray luminosity (between 0.5 and 10 keV) of a few times of $10^{33}\ \rm erg\ s^{-1}$, 
since at this X-ray luminosity the calculated $\dot M$ with the model of ADAF accretion can be 
high enough, e.g., more than two orders of magnitude higher than that of calculated with the formula of 
$\dot M=L_{\rm X}{R_{*}\over {GM}}$ for taking $f_{\rm th}=0.01$, to drive a fraction of the matter in the 
accretion flow to be channelled onto the surface of the NS forming the X-ray pulsation.
%---------
A brief summary on the ADAF model around a NS and the constraints to the value of $f_{\rm th}$ in 
\citet[][]{Qiao2020a,Qiao2020b} are introduced in Section 2. The results are shown in Section 3.
The discussions are in Section 4 and the conclusions are in Section 5.

\section{A summary on Qiao \& Liu 2020a,b}
The structure and the corresponding emergent spectra of the ADAF around a NS are strictly investigated 
within the framework of the self-similar solution of the ADAF \citep[][]{Qiao2018b}. 
In \citet[][]{Qiao2020a,Qiao2020b}, we update the code with the effect of the NS spin considered compared 
with that of in \citet[][]{Qiao2018b}. 
In our model, there are seven parameters, i.e., the NS mass $m$ ($m=M/M_{\odot}$), NS radius $R_{*}$,
NS spin frequency $\nu_{\rm NS}$, mass accretion rate $\dot m$ ($\dot m=\dot M/\dot M_{\rm Edd}$), 
as well as the viscosity parameter $\alpha$, and the magnetic parameter 
$\beta$ [with magnetic pressure $p_{\rm m}={B^2/{8\pi}}=(1-\beta)p_{\rm tot}$, $p_{\rm tot}=p_{\rm gas}+p_{\rm m}$]
for describing the microphysics of the ADAF. The last parameter is, $f_{\rm th}$, describing the fraction of 
the ADAF energy released at the surface of the NS as thermal emission to be scattered in the ADAF to
cool the ADAF itself, which consequently controls the feedback between the ADAF and the NS.  
We always take $m=1.4$, and $R_{*}$ in the range of 10-12.5 km \citep[][]{Degenaar2018b}
[$R_{*}=12.5$ km in \citet[][]{Qiao2020a}, and $R_{*}=10$ km in \citet[][]{Qiao2020b}]. 
In general, it has been proven that the effect of the NS spin frequency $\nu_{\rm NS}$ on the structure and 
the emergent spectra of the ADAF around a NS is very little, and nearly can be 
neglected [see Figure 8 of \citet[][]{Qiao2020a} for taking $\nu_{\rm NS}=0, 200, 500, 700$ Hz respectively]. 
So we fix $\nu_{\rm NS}=0$ Hz in \citet[][]{Qiao2020a,Qiao2020b}. The magnetic field in ADAF is very weak as 
suggested by the magnetohydrodynamic simulations \citep[][for review]{Yuan2014}. We fix $\beta=0.95$ in 
\citet[][]{Qiao2020a,Qiao2020b}. 
 
%----------
The X-ray spectra of NS-LMXBs in the low-level accretion regime 
($L_{\rm 0.5-10\rm keV} \lesssim 10^{36} \rm \ erg \ s^{-1}$ ) can be described by a single power-law model, 
or a two-component model, i.e., a thermal soft X-ray component plus a power-law component. 
In general, if the $\it Swift$ X-ray data are used, the spectral fitting with a single power-law model can return 
an accepted fit. And if the high-quality $\it XMM-Newton$ X-ray data are used, the spectra fitting with a 
two-component model can significantly improve the fitting results in some X-ray luminosity range , e.g., 
in the range of $L_{\rm 0.5-10\rm keV} \sim 10^{34}-10^{35} \rm \ erg \ s^{-1}$ \citep[e.g.][]{Wijnands2015}.
%----------
In \citet[][]{Qiao2020a}, we test the effect of $\alpha$
and $f_{\rm th}$ on the X-ray spectra between 0.5 and 10 keV, and explain the fractional contribution of the 
power-law component $\eta$ ($\eta\equiv L^{\rm power\ law}_{\rm 0.5-10\rm keV}/L_{\rm 0.5-10\rm keV}$) (with 
the spectra fitted with the two-component model) as a function of the $L_{\rm 0.5-10keV}$ for a sample of 
non-pulsating NS-LMXBs in a wide range from $L_{\rm 0.5-10keV}\sim 10^{32}-10^{36}\ \rm erg\ s^{-1}$.
%----------
Observationally, there is a positive correlation between $\eta$ and $L_{\rm 0.5-10keV}$ for 
$L_{\rm 0.5-10\rm keV} \gtrsim$ a few times of $10^{33} \rm \ erg \ s^{-1}$, and an anticorrelation between 
$\eta$ and $L_{\rm 0.5-10keV}$ for $L_{\rm 0.5-10\rm keV} \lesssim$ a few times of $10^{33} \rm \ erg \ s^{-1}$.
By comparing with the observed correlation (both the positive correlation and the anticorrelation) between 
$\eta$ and $L_{\rm 0.5-10keV}$, it is found that the effect of $\alpha$ on the correlation between 
$\eta$ and $L_{\rm 0.5-10keV}$ is very little, and nearly can be neglected. Meanwhile, it is found that the 
correlation between $\eta$ and $L_{\rm 0.5-10keV}$ can be well matched by adjusting  the value of $f_{\rm th}$. 
The value of $f_{\rm th}$ is constrained to be less than 0.1. Especially, $f_{\rm th}=0.01$ is more preferred 
[see Figure 7 of \citet[][]{Qiao2020a}].
%--------------------------

Further, in \citet[][]{Qiao2020b}, based on the sample of non-pulsating NS-LMXBs in \citet[][]{Wijnands2015}, 
and adding some more non-pulsating NS-LMXBs from \citet[][]{Parikh2017} and \citet[][]{Beri2019}, we explain the 
anticorrelation between the X-ray photon index $\Gamma$ (obtained by fitting the X-ray spectra between 0.5 and 10 keV 
with a single power law) and $L_{\rm 0.5-10keV}$, i.e., the softening of the X-ray spectra with decreasing 
$L_{\rm 0.5-10keV}$, in the range of $L_{\rm 0.5-10keV}\sim 10^{34}-10^{36}\ \rm erg\ s^{-1}$ 
by adjusting the value of $f_{\rm th}$.
%-------
Moreover, it is shown that a fraction of the sources in \citet[][]{Qiao2020b} are once reported to be fitted with 
the two-component model (with $\it XMM-Newton$ X-ray data), i.e., a thermal soft X-ray component plus a power-law 
component. 
%-------
Combining the explanations for the anticorrelation between the X-ray photon index $\Gamma$ and 
$L_{\rm 0.5-10keV}$ with the X-ray spectra analyzed with the single power-law model, and the positive correlation 
between $\eta$ and $L_{\rm 0.5-10keV}$ with the two-component model for a fraction of the sources in the sample, 
we conclude that in the range of $L_{\rm 0.5-10keV}\sim 10^{34}-10^{35}\ \rm erg\ s^{-1}$,
the softening of the X-ray spectra is due to the increase of the thermal soft X-ray component,
while in the range of $L_{\rm 0.5-10keV}\sim 10^{35}-10^{36}\ \rm erg\ s^{-1}$,  
the softening of the X-ray spectra is probably due to the evolution of the power-law component itself.
%-------
As a summary, in the study above for explaining the anticorrelation between $\Gamma$ and $L_{\rm 0.5-10keV}$, 
it has been shown that the value of $f_{\rm th}$ can be constrained to be less than 0.1, and is very probably 
to be much smaller values, i.e., $\sim 0.003-0.005$ \citep[][]{Qiao2020b}. 

In the following, fixing $m=1.4$, $R_{*}=12.5$ km, $\nu_{\rm NS}=500$ Hz (see Section \ref {s:pulsation} for 
discussions), $\alpha=0.3$ and $\beta=0.95$, we investigate the radiative 
efficiency of NSs with an ADAF accretion for different $\dot m$ by taking two typical values of 
$f_{\rm th}$, i.e., $f_{\rm th}=0.1$ and $f_{\rm th}=0.01$ respectively.

\section{Results}
\subsection{Numerical results}\label{s:results}
\begin{figure*}
\includegraphics[width=85mm,height=60mm,angle=0.0]{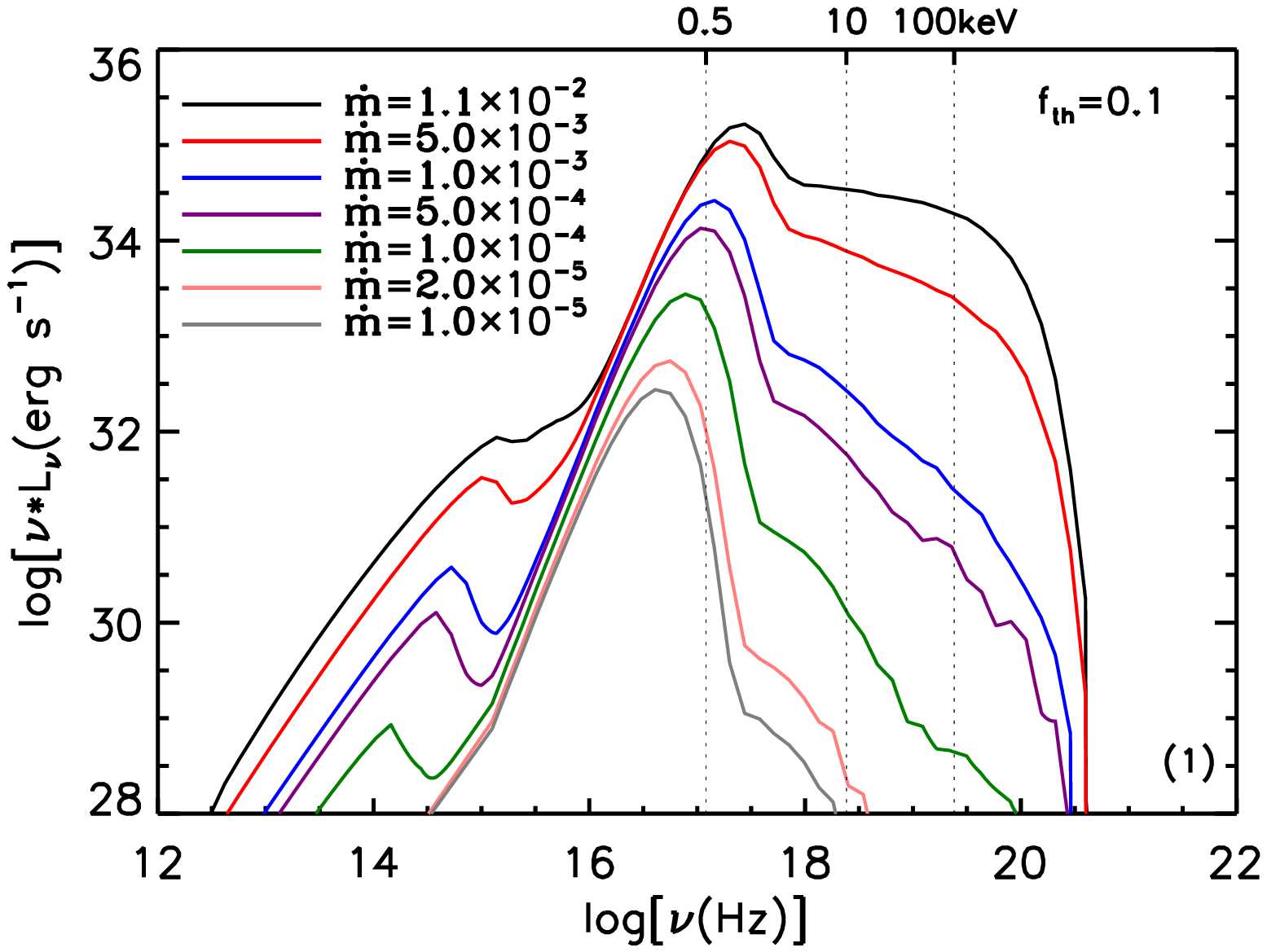}
\includegraphics[width=85mm,height=60mm,angle=0.0]{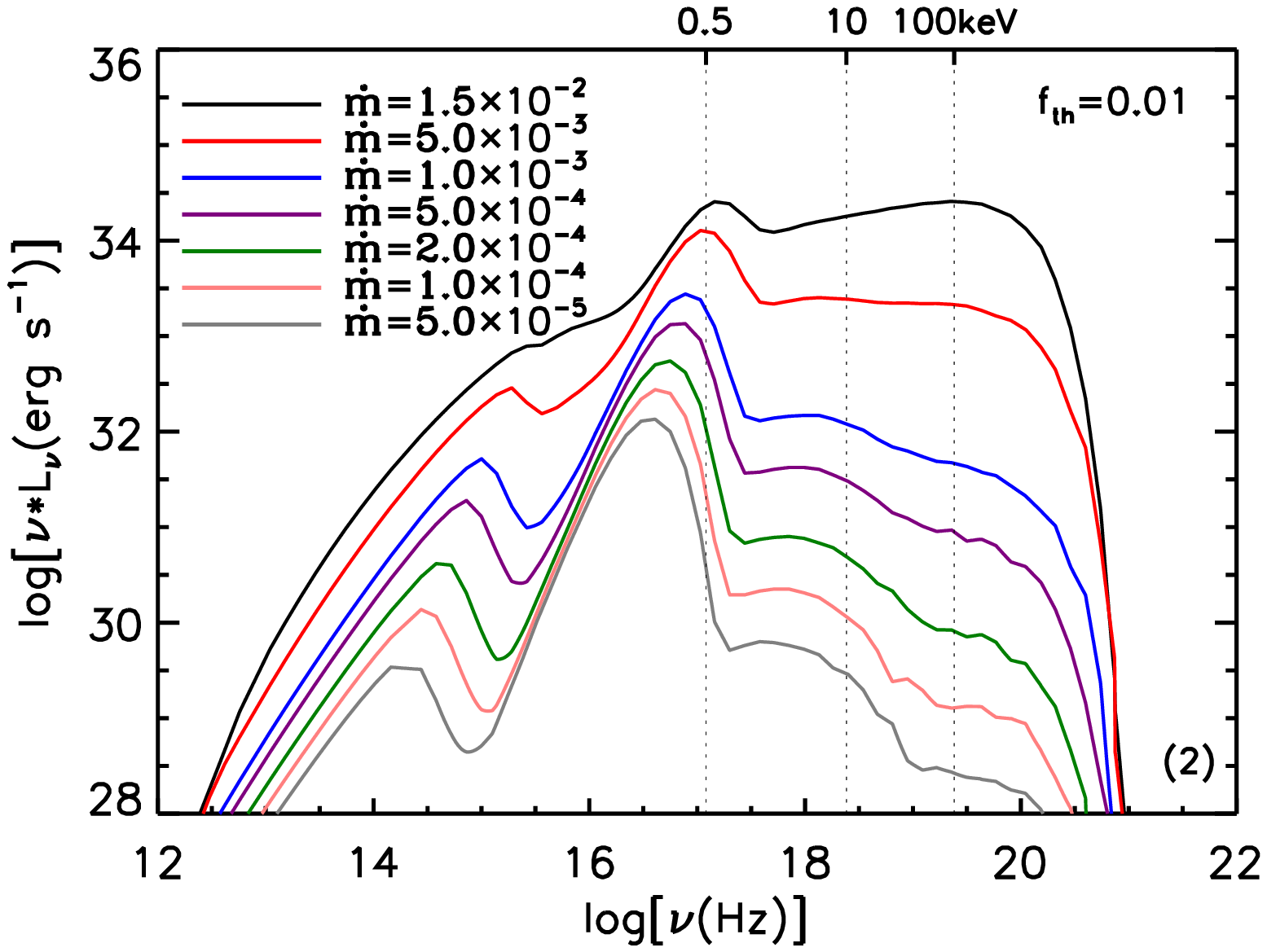}
\caption{\label{f:sp} Panel (1): emergent spectra of NSs with an ADAF accretion for different $\dot m$ 
with $f_{\rm th}=0.1$. Panel (2): emergent spectra of NSs with an ADAF accretion for different $\dot m$ 
with $f_{\rm th}=0.01$. 
%In all the calculations, we take $m=1.4$, $R_{*}=12.5\rm km$, $\alpha=0.3$, and $\beta=0.95$ respectively. 
}
\end{figure*}

\begin{figure*}
\includegraphics[width=85mm,height=60mm,angle=0.0]{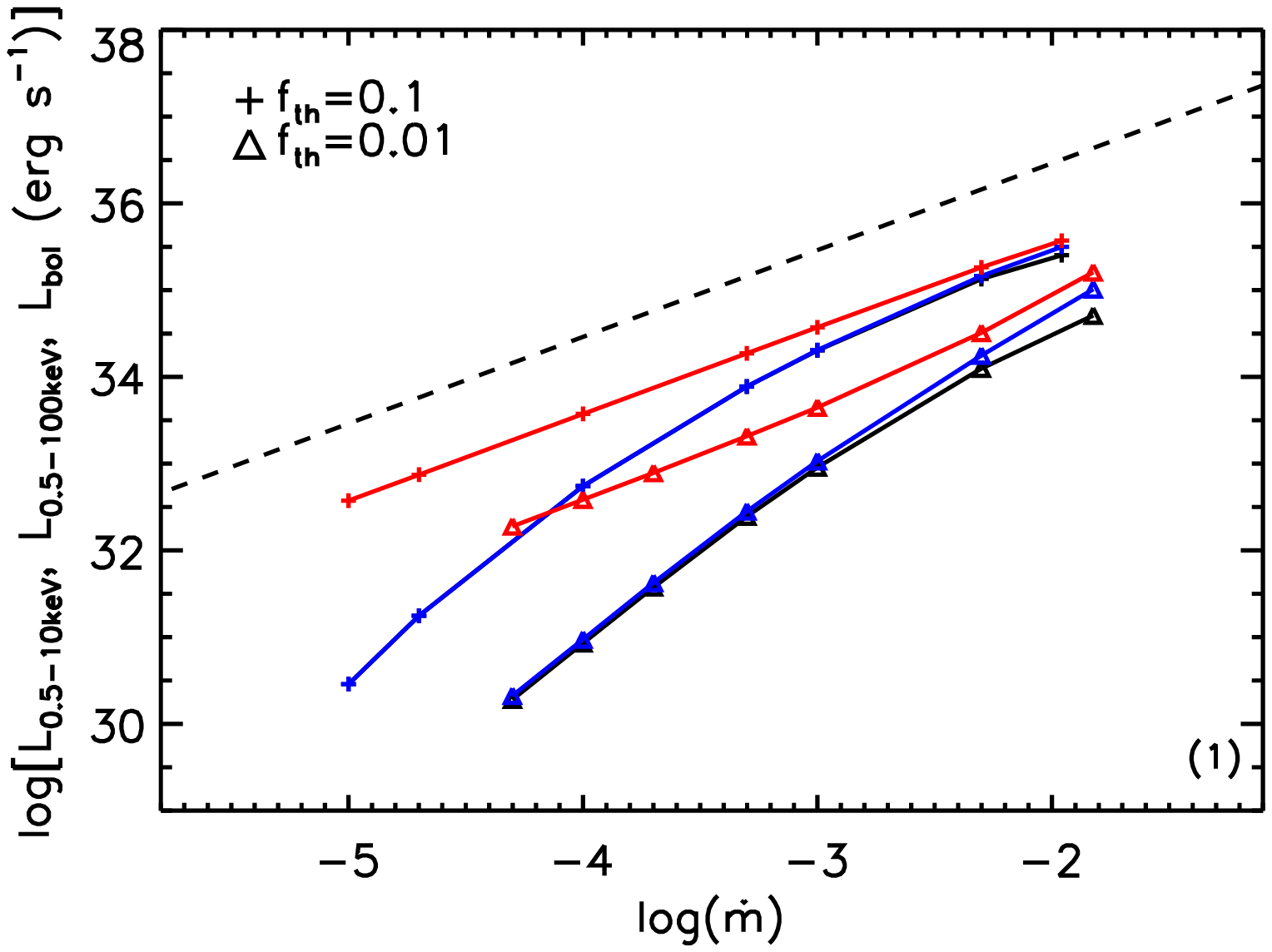}
\includegraphics[width=85mm,height=60mm,angle=0.0]{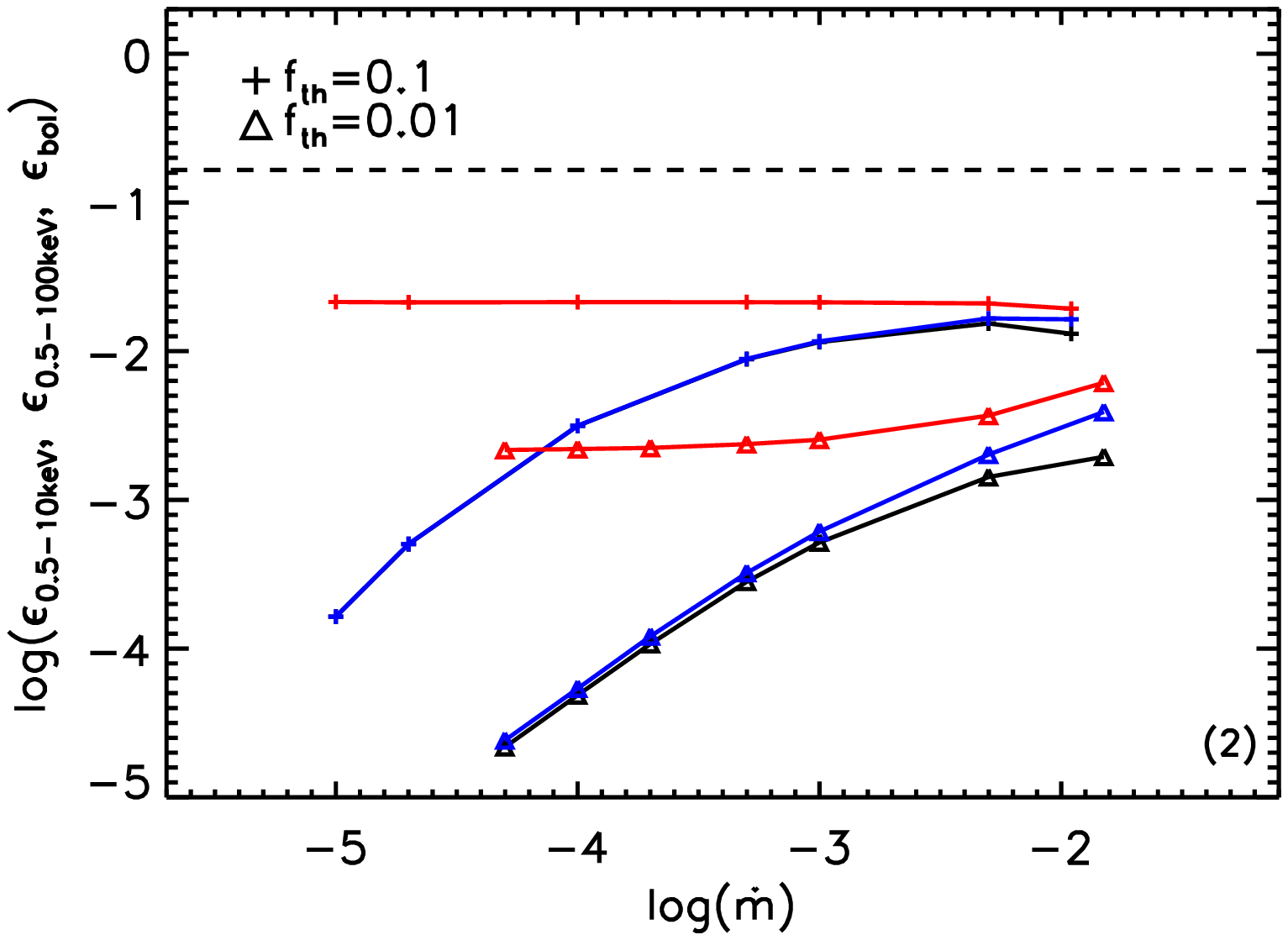}
\caption{\label{f:effi} Panel (1): X-ray luminosity $L_{\rm 0.5-10keV}$, $L_{\rm 0.5-100keV}$ and 
bolometric luminosity $L_{\rm bol}$ as a function of $\dot m$ respectively. 
%------
The symbols of black `+', blue `+', and red `+' refer to $L_{\rm 0.5-10keV}$, $L_{\rm 0.5-100keV}$, and $L_{\rm bol}$ 
respectively from our model of ADAF accretion with $f_{\rm th}=0.1$. 
%------
The symbols of black `$\triangle$', blue `$\triangle$', and red `$\triangle$' 
refer to $L_{\rm 0.5-10keV}$, $L_{\rm 0.5-100keV}$, and $L_{\rm bol}$ respectively from our model of ADAF 
accretion with $f_{\rm th}=0.01$.
%------
The dashed line refers to the luminosity calculated with the formula of $L={\dot M GM\over R_{*}}$. 
%------------------
Panel (2): radiative efficiency $\epsilon_{\rm 0.5-10keV}$, $\epsilon_{\rm 0.5-100keV}$ 
and $\epsilon_{\rm bol}$ as a function of $\dot m$. 
%------
The symbols of black `+', blue `+', and red `+' refer to $\epsilon_{\rm 0.5-10keV}$, $\epsilon_{\rm 0.5-100keV}$ 
and $\epsilon_{\rm bol}$ respectively from our model of ADAF accretion with $f_{\rm th}=0.1$. 
%------
The symbols of black `$\triangle$', blue `$\triangle$', and red `$\triangle$' refer to $\epsilon_{\rm 0.5-10keV}$, 
$\epsilon_{\rm 0.5-100keV}$ and $\epsilon_{\rm bol}$ respectively from our model of ADAF accretion with 
$f_{\rm th}=0.01$. 
%------
The dashed line refers to the radiative efficiency calculated with the formula of 
$\epsilon={\dot M GM\over R_{*}}/{\dot Mc^2}$. 
%In all the calculations, we take $m=1.4$, $R_{*}=12.5\rm km$, $\alpha=0.3$, and $\beta=0.95$ respectively. 
}
\end{figure*}

We plot the emergent spectra of NSs with an ADAF accretion for different $\dot m$ with $f_{\rm th}=0.1$ in 
panel (1) of Fig. \ref{f:sp}, and with $f_{\rm th}=0.01$ in panel (2) of Fig. \ref{f:sp}.
%[the emergent spectra in panel (1) is similar to that of in panel (2) of Fig. (5) of \citet[][]{Qiao2020a} 
%except that a new spectrum for $\dot m=1.1\times 10^{-2}$ is added for completeness; the emergent spectra in 
%panel (2) is similar to that of in panel (4) of Fig. (5) of \citet[][]{Qiao2020a} except that a new spectrum for 
%$\dot m=5.0\times 10^{-5}$ is added for completeness.]
%---------
Based on the emergent spectra in panel (1) of Fig. \ref{f:sp}, we calculate three quantities, i.e., the X-ray 
luminosity between 0.5 and 10 keV $L_{\rm 0.5-10keV}$, the X-ray luminosity between 0.5 and 100 keV $L_{\rm 0.5-100keV}$, 
and the bolometric luminosity $L_{\rm bol}$ for different $\dot m$ with $f_{\rm th}=0.1$.
%---------
Based on the emergent spectra in panel (2) of Fig. \ref{f:sp}, a similar calculation is done 
for $L_{\rm 0.5-10keV}$, $L_{\rm 0.5-100keV}$, and $L_{\rm bol}$ for different $\dot m$ with $f_{\rm th}=0.01$.
%---------
%---------
In panel (1) of Fig. \ref{f:effi}, we plot $L_{\rm 0.5-10keV}$, $L_{\rm 0.5-100keV}$, and $L_{\rm bol}$ as 
a function of $\dot m$ for $f_{\rm}=0.1$ and $f_{\rm}=0.01$ respectively.
%---------
Specifically, for $f_{\rm}=0.1$, it can be seen that, all the three quantities $L_{\rm 0.5-10keV}$, 
$L_{\rm 0.5-100keV}$, and $L_{\rm bol}$ decrease with decreasing $\dot m$.
Meanwhile, it can be seen that, for $f_{\rm}=0.1$, $L_{\rm 0.5-10keV}$ as a function of $\dot m$ is nearly 
overlapped with $L_{\rm 0.5-100keV}$ as a function of $\dot m$.
This is because all the X-ray spectra are very soft for different $\dot m$ with $f_{\rm th}=0.1$, the 
value of $L_{\rm 0.5-10keV}$ and $L_{\rm 0.5-100keV}$ is nearly same for a fixed $\dot m$.
It also easy to see that, for $f_{\rm}=0.1$, the value of $L_{\rm bol}$ is always greater than 
$L_{\rm 0.5-10keV}$ (or $L_{\rm 0.5-100keV}$). Meanwhile, the separation between $L_{\rm bol}$ and 
$L_{\rm 0.5-10keV}$ (or $L_{\rm 0.5-100keV}$) becomes larger and larger with decreasing $\dot m$.
%---------
In general, the trends of $L_{\rm 0.5-10keV}$, $L_{\rm 0.5-100keV}$, and $L_{\rm bol}$ as a function of 
$\dot m$ for $f_{\rm}=0.01$ are similar to that of for $f_{\rm}=0.1$ respectively.
However, the values of $L_{\rm 0.5-10keV}$, $L_{\rm 0.5-100keV}$, and $L_{\rm bol}$ for $f_{\rm}=0.01$ 
are systematically lower than that of for $f_{\rm}=0.1$ for roughly one order of magnitude or more for a fixed 
$\dot m$ respectively. 
%---------
We further plot the formula $L={\dot M GM\over R_{*}}$ (note: $\dot M=\dot m \dot M_{\rm Edd}$, $M=m M_{\odot}$) 
as a comparison, one can refer to the dashed line in panel (1) of Fig. \ref{f:effi} for clarity. It can be seen that 
all the three luminosities, i.e., $L_{\rm 0.5-10keV}$, $L_{\rm 0.5-100keV}$ and $L_{\rm bol}$, are lower than the 
luminosity calculated with the formula of $L={\dot M GM\over R_{*}}$ for a fixed $\dot M$ (or $\dot m$). 

We define three quantities for the radiative efficiency in different bands, i.e., 
\begin{eqnarray}\label{e:bol} 
\epsilon_{\rm 0.5-10keV}=L_{\rm 0.5-10keV}/{\dot M c^{2}}, 
\end{eqnarray}
\begin{eqnarray}\label{e:bol} 
\epsilon_{\rm 0.5-100keV}=L_{\rm 0.5-100keV}/{\dot M c^{2}}, 
\end{eqnarray}
\begin{eqnarray}\label{e:bol} 
\epsilon_{\rm bol}=L_{\rm bol}/{\dot M c^{2}}.
\end{eqnarray}
In panel (2) of Fig. \ref{f:effi}, we plot $\epsilon_{\rm 0.5-10keV}$, $\epsilon_{\rm 0.5-100keV}$ and 
$\epsilon_{\rm bol}$ as a function of $\dot m$ with $f_{\rm}=0.1$, and $f_{\rm}=0.01$ respectively.
%-------------------------
Specifically, for $f_{\rm th}=0.1$, $\epsilon_{\rm bol}$ is nearly a constant for different $\dot m$ with 
$\epsilon_{\rm bol}\sim 0.02$. $\epsilon_{\rm 0.5-10keV}$ as a function of $\dot m$ is nearly overlapped 
with $\epsilon_{\rm 0.5-100keV}$ as a function of $\dot m$. 
For $f_{\rm th}=0.1$, $\epsilon_{\rm 0.5-10keV}$ decreases from $\sim 0.013$ to $\sim 1.6\times 10^{-4}$ for $\dot m$ 
decreasing from $1.1\times 10^{-2}$ to $1.0\times 10^{-5}$, and $\epsilon_{\rm 0.5-100keV}$ decreases 
from $\sim 0.016$ to $\sim 1.6\times 10^{-4}$ for $\dot m$ decreasing 
from $1.1\times 10^{-2}$ to $1.0\times 10^{-5}$.
%---------
In general, the trends of $\epsilon_{\rm 0.5-10keV}$, $\epsilon_{\rm 0.5-100keV}$ and $\epsilon_{\rm bol}$ 
as a function of $\dot m$ for $f_{\rm th}=0.01$ are similar to that of for  $f_{\rm th}=0.1$.
For $f_{\rm th}=0.01$, $\epsilon_{\rm bol}$ is also nearly a constant, decreasing slightly with decreasing 
$\dot m$, i.e., $\epsilon_{\rm bol}$ decreasing from $\sim 0.006$ to $\sim 0.002$ for 
$\dot m$ decreasing from $1.5\times 10^{-2}$ to $5.0\times 10^{-5}$.
%---------
For $f_{\rm th}=0.01$, $\epsilon_{\rm 0.5-10keV}$ as a function of $\dot m$ is also nearly overlapped with 
$\epsilon_{\rm 0.5-100keV}$ as a function of $\dot m$. 
Specifically, $\epsilon_{\rm 0.5-10keV}$ decreases from $\sim 0.002$ to $\sim 2.2\times 10^{-5}$ for $\dot m$ 
decreasing from $1.5\times 10^{-2}$ to $5.0\times 10^{-5}$.
$\epsilon_{\rm 0.5-100keV}$ decreases from $\sim 0.004$ to $\sim 2.4\times 10^{-5}$ for $\dot m$ 
decreasing from $1.5\times 10^{-2}$ to $5.0\times 10^{-5}$.
Also as a comparison, we plot the radiative efficiency $\epsilon$ 
calculated with the formula of $\epsilon \sim {\dot M GM\over R_{*}}/{\dot M c^2}$,
see the dashed line in panel (2) of Fig. \ref{f:effi}. The value of $\epsilon$  
is $\sim 0.2$, which is roughly one order of magnitude higher than $\epsilon_{\rm bol}$ for $f_{\rm}=0.1$ 
and two orders of magnitude higher than $\epsilon_{\rm bol}$ for $f_{\rm}=0.01$. 

\begin{figure}
\includegraphics[width=85mm,height=60mm,angle=0.0]{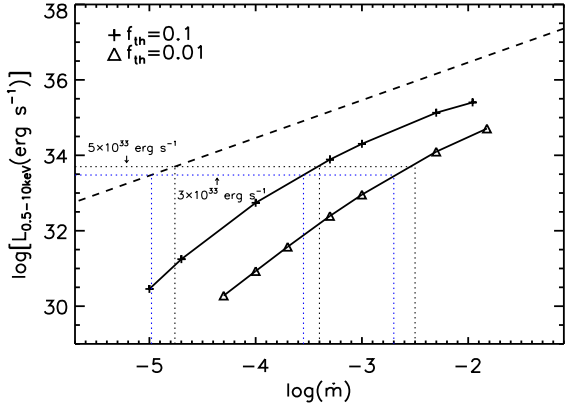}
\caption{\label{f:mdot} X-ray luminosity $L_{\rm 0.5-10keV}$ as a function of $\dot m$.
The symbols of `+' and `$\triangle$' refer to  $L_{\rm 0.5-10keV}$ from our model of ADAF accretion with 
$f_{\rm th}=0.1$ and $f_{\rm th}=0.01$ respectively. The dashed line refers to the luminosity calculated with 
the formula of $L={\dot M GM\over R_{*}}$. The horizontal black-dotted line refers to the X-ray luminosity of 
$L_{\rm 0.5-10keV}=5\times 10^{33}\ \rm erg \ s^{-1}$, and the horizontal blue-dotted line refers to 
the X-ray luminosity of $L_{\rm 0.5-10keV}=3\times 10^{33}\ \rm erg \ s^{-1}$.
%-----------
The vertical black-dotted lines from left to right refer to the value of $\dot m$ of the crossing points 
between $L_{\rm 0.5-10keV}=5\times 10^{33}\ \rm erg \ s^{-1}$ and the formula of $L={\dot M GM\over R_{*}}$,
our model of ADAF accretion with $f_{\rm th}=0.1$, and our model of ADAF accretion with 
$f_{\rm th}=0.01$ respectively.
%-----------
The vertical blue-dotted lines from left to right refer to the value of $\dot m$ of the crossing points between   
$L_{\rm 0.5-10keV}=3\times 10^{33}\ \rm erg \ s^{-1}$ and the formula of $L={\dot M GM\over R_{*}}$,
our model of ADAF accretion with $f_{\rm th}=0.1$, and our model of ADAF accretion with 
$f_{\rm th}=0.01$ respectively.  }
\end{figure}

In summary, based on our study above for taking $f_{\rm th}=0.1$ and $f_{\rm th}=0.01$,
the predicted luminosity, i.e., $L_{\rm 0.5-10keV}$, $L_{\rm 0.5-100keV}$ and $L_{\rm bol}$ from our model of 
ADAF accretion, are all lower than the luminosity predicted by the formula of $L={\dot M GM\over R_{*}}$ for a 
fixed $\dot M$ (or $\dot m$). This in turn means that, given a value of   
$L_{\rm 0.5-10keV}$, $L_{\rm 0.5-100keV}$ or $L_{\rm bol}$, the obtained $\dot M$ (or $\dot m$) from our model 
of ADAF accretion is greater than that of calculated with the formula of $\dot M=L{R_{*}\over {GM}}$ ($L$ can 
be $L_{\rm 0.5-10keV}$, $L_{\rm 0.5-100keV}$ or $L_{\rm bol}$). 
%------
Here we just take two X-ray luminosities between 0.5 and 10 keV, i.e.,  
$L_{\rm 0.5-10keV}=3.0\times 10^{33}\ \rm erg\ s^{-1}$ and 
$L_{\rm 0.5-10keV}=5.0\times 10^{33}\ \rm erg\ s^{-1}$\footnote{We take these two X-ray luminosities since
the X-ray pulsations have been confirmed in some NS-LMXBs in these luminosities, as will be discussed in 
Section \ref {s:pulsation}.} as examples for calculating $\dot M$.
One can refer to Fig. \ref{f:mdot} for the illustrations and Table \ref{t:mdot} for the detailed numerical 
results. 
%------
Specifically, for $L_{\rm 0.5-10keV}=3.0\times 10^{33}\ \rm erg\ s^{-1}$, the mass accretion rate $\dot M_{0}$ 
calculated with the formula of $\dot M_{0}=L_{\rm 0.5-10keV}{R_{*}\over {GM}}$
is $2.02\times 10^{13}\ \rm g \ s^{-1}$, which is $\sim 27$ times less than the mass accretion rate 
$\dot M_{\rm 0.1}$ calculated with our model of ADAF accretion for taking $f_{\rm th}=0.1$, 
and is $\sim 192$ times less than the mass accretion rate $\dot M_{\rm 0.01}$ calculated with our model of 
ADAF accretion for taking $f_{\rm th}=0.01$. 
%------
For $L_{\rm 0.5-10keV}=5.0\times 10^{33}\ \rm erg\ s^{-1}$, $\dot M_{0}$ calculated
with the formula of $\dot M_{0}=L_{\rm 0.5-10keV}{R_{*}\over {GM}}$ is $3.36\times 10^{13}\ \rm g \ s^{-1}$, 
which is $\sim 23$ times less than $\dot M_{\rm 0.1}$ calculated with our model of ADAF accretion for taking 
$f_{\rm th}=0.1$, and is $\sim 183$ times less than $\dot M_{\rm 0.01}$ calculated with our model of ADAF accretion 
for taking $f_{\rm th}=0.01$. 
%------
The relatively higher $\dot M$ calculated from our model of ADAF accretion has very clear physical meanings, 
as will be discussed in Section \ref {s:pulsation}.

\begin{table*}
\caption{Mass accretion rate obtained for a given X-ray luminosity $L_{\rm 0.5-10keV}$ based on the curves in  
Fig. \ref{f:mdot}. Specifically, $\dot m_{0}$ and $\dot M_{0}$ are obtained with the formula of 
$\dot M_{0}=L_{\rm 0.5-10keV}{R_{*}\over {GM}}$. 
$\dot m_{0}$ and $\dot M_{0}$ are in units of $\dot M_{\rm Edd}$ and $\rm g\ s^{-1}$ respectively.
%--------------
$\dot m_{0.1}$ and $\dot M_{0.1}$ are obtained from our model results
of NSs with an ADAF accretion for $f_{\rm th}=0.1$. 
$\dot m_{0.1}$ and $\dot M_{0.1}$ are in units of $\dot M_{\rm Edd}$ and $\rm g\ s^{-1}$ respectively.
%--------------
$\dot m_{0.01}$ and $\dot M_{0.01}$ are obtained from our model results 
of NSs with an ADAF accretion for $f_{\rm th}=0.01$. 
$\dot m_{0.01}$ and $\dot M_{0.01}$ are in units of $\dot M_{\rm Edd}$ and $\rm g\ s^{-1}$ respectively.
%--------------
%In all the calculations, we take $m=1.4$, $R_{*}=12.5\rm km$, $\alpha=0.3$, and $\beta=0.95$ respectively.     
}
\centering
\begin{tabular}{ccccccc}
\hline
$L_{\rm 0.5-10keV}\ (\rm \ erg \ s^{-1})$ & $\dot m_{0}$ ($\dot M_{\rm Edd}$) & $\dot M_{0}$ ($\rm g\ s^{-1}$) & 
$\dot m_{0.1}$  ($\dot M_{\rm Edd}$)  &  $\dot M_{0.1}$   ($\rm g\ s^{-1}$)   & 
$\dot m_{0.01}$ ($\dot M_{\rm Edd}$)  &  $\dot M_{0.01}$  ($\rm g\ s^{-1}$)  \\
\hline
$3.0\times 10^{33}$   &  $1.04\times 10^{-5}$  &  $2.02\times 10^{13}$ &
$2.82\times 10^{-4}$  &  $5.48\times 10^{14}$  &
$2.00\times 10^{-3}$  &  $3.88\times 10^{15}$  \\
$5.0\times 10^{33}$   &  $1.73\times 10^{-5}$  &  $3.36\times 10^{13}$ & 
$3.98\times 10^{-4}$  &  $7.75\times 10^{14}$   &
$3.16\times 10^{-3}$  &  $6.15\times 10^{15}$  \\ 
\hline
\end{tabular}
\\
\label{t:mdot}
\end{table*}

\subsection{Possible applications for explaining the formation of the millisecond X-ray pulsation 
at the X-ray luminosity of a few times of $10^{33}\ \rm erg\ s^{-1}$}\label{s:pulsation}
Recently, the millisecond X-ray pulsations have been observed in several NS X-ray sources, such as, the transitional 
millisecond pulsar (tMSP) PSR J1023+0038 \citep[][]{Archibald2015} and  XSS J12270-4859 \citep[][]{Papitto2015} as 
they are in the accretion-powered LMXB state, as well as the X-ray transient IGR J17379-3747 \citep[][]{Bult2019}, 
at a lower X-ray luminosity (between 0.5 and 10 keV) of a few times of $10^{33}\ \rm erg\ s^{-1}$. 
This challenges the traditional accretion disc theory for the formation of the X-ray pulsation at such 
a low X-ray luminosity, since in general at this low X-ray luminosity (if the mass accretion rate calculated 
with formula of $\dot M=L_{\rm X}{R_{*}\over {GM}}$), the corotation radius of the NS accreting system is 
less than the magnetospheric radius. In this case, the `propeller' effect may work, expelling 
(a fraction of the) matter in the accretion flow to be leaving away from the NS \citep[][]{Illarionov1975}. 
%---------------
In this paper, we suggest that, if applicable, our model of ADAF accretion may be applied to explain the observed 
millisecond X-ray pulsation at the X-ray luminosity of a few times of $10^{33}\ \rm erg\ s^{-1}$. 
This is because at this X-ray luminosity, $\dot M$ calculated from our model of ADAF accretion for taking 
an appropriate value of $f_{\rm th}$, such as $f_{\rm th}=0.01$, 
can be more than two orders of magnitude higher than that of calculated with the formula of 
$\dot M=L_{\rm X}{R_{*}\over {GM}}$ to make {the magnetospheric radius less than
the corotation radius. In this case, as the accretion flow moves inward, if the radius is less than 
the magnetospheric radius, (a fraction of) the matter in the accretion flow will be magnetically channelled 
onto the surface of the NS, leading to the formation of the X-ray pulsation.} 

%---------------
For clarity, we list the expression of the corotation radius $R_{\rm c}$ and the magnetospheric radius 
$R_{\rm m}$ respectively as follows. The corotation radius $R_{\rm c}$ is expressed as,
%---------------
\begin{eqnarray}\label{e:rc} 
R_{\rm c} & = & \biggl({GM\over {4\pi^2\nu_{\rm NS}^2}}\biggr)^{1/3}  \nonumber \\
    & \simeq & 26.6\ {\rm km}\ \biggl({M\over {1.4M_{\odot}}}\biggl)^{1/3}\biggl({\nu_{\rm NS}\over {500\ \rm Hz}}\biggr)^{-2/3},
\end{eqnarray}
where $M$ is the NS mass, and $\nu_{\rm NS}$ is the NS spin frequency.
The magnetospheric radius $R_{\rm m}$ is expressed as \citep[][]{Spruit1993,DAngelo2010,DAngelo2015},
%--------
\begin{eqnarray}\label{e:rm} 
R_{\rm m} & = & \biggl({\eta \mu^2 \over {4\Omega_{\rm NS} \dot M}}\biggr)^{1/5}  \nonumber \\
& \simeq & 24\ {\rm km}\ \eta^{1/5}\biggl({B_{*}\over {10^8\ \rm G}}\biggr)^{2/5}\biggl({R_{*}\over {10\ \rm km}}\biggr)^{6/5} \nonumber \\
&   & \times\ \biggl({\nu_{\rm NS}\over {500\ \rm Hz}}\biggr)^{-1/5}\biggl({\dot M\over {10^{16}\ \rm g\ s^{-1}}}\biggr)^{-1/5},
\end{eqnarray}
where $\mu=B_{*}R_{*}^3$ is magnetic dipole moment (with $B_{*}$ being the magnetic field at
the surface of the NS, $R_{*}$ being the NS radius), $\Omega_{\rm NS}$ is the rotational angular velocity of the NS 
(with $\Omega_{\rm NS}=2\pi\nu_{\rm NS}$), $\eta\leq 1$ is the dimensionless parameter describing the strength of 
the toroidal magnetic field induced by the relative rotation between the accretion flow and dipolar magnetic field, 
and $\dot M$ is the mass accretion rate in units of $\rm g\ s^{-1}$. We investigate the relation between $R_{\rm c}$ 
and $R_{\rm m}$ for PSR J1023+0038, XSS J12270-4859 and IGR J17379-3747 respectively as follows. 

{\bf PSR J1023+0038:} the millisecond X-ray pulsation of PSR J1023+0038 has been discovered at the X-ray 
luminosity of $L_{\rm 0.5-10keV}\sim 3.0\times 10^{33} \ \rm erg \ s^{-1}$. 
The spin frequency of PSR J1023+0038 is $\nu_{\rm NS}=592$ Hz \citep[][]{Archibald2015}. 
As we can see from Table \ref{t:mdot}, at the X-ray luminosity of 
$L_{\rm 0.5-10keV}\sim 3.0\times 10^{33} \ \rm erg \ s^{-1}$, the mass accretion rate $\dot M_{0}$ calculated 
with the formula of $\dot M_{0}=L_{\rm 0.5-10keV}{R_{*}\over {GM}}$ is $2.02\times 10^{13}\ \rm g \ s^{-1}$.
At this X-ray luminosity, the mass accretion rate $\dot M_{0.1}$ calculated with our model of ADAF accretion 
for $f_{\rm th}=0.1$ is $5.48\times 10^{14}\ \rm g \ s^{-1}$, 
and mass accretion rate $\dot M_{0.01}$ calculated with our model of ADAF accretion 
for $f_{\rm th}=0.01$ is $3.88\times 10^{15}\ \rm g \ s^{-1}$. 
%----------
If we assume $M=1.4M_{\odot}$, $R_{*}=12.5\ \rm km$ as we take in Section \ref{s:results} of this paper, 
a typical value of the magnetic field at the surface of the NS $B_{*}=10^8\ \rm G$ and $\eta=0.1$, 
according to equation (\ref{e:rc}), $R_{\rm c}\approx 23.8\ \rm km$, and according to equation (\ref{e:rm}),
$R_{\rm m0}\approx 66.2\ \rm km$, $R_{\rm m0.1}\approx 34.2\ \rm km$, and $R_{\rm m0.01}\approx 23.1\ \rm km$  
(with $R_{\rm m0}$, $R_{\rm m0.1}$ and $R_{\rm m0.01}$ being the magnetospheric radii calculated by taking the 
mass accretion rate as $\dot M_{0}$, $\dot M_{0.1}$ and $\dot M_{0.01}$ respectively). 
%----------
It can be seen that, if the mass accretion rate, i.e. $\dot M_{0}$, is calculated with the formula of   
$\dot M_{0}=L_{\rm 0.5-10keV}{R_{*}\over {GM}}$, $R_{\rm m0}>R_{\rm c}$, theoretically, in this case the pulsation 
cannot be formed. If the mass accretion rate, i.e. $\dot M_{0.1}$, is calculated with our model of ADAF accretion
for $f_{\rm th}=0.1$, $R_{\rm m0.1}>R_{\rm c}$, theoretically, the pulsation also cannot be formed.
While, if the mass accretion rate, i.e. $\dot M_{0.01}$, is calculated with our model of ADAF accretion
for $f_{\rm th}=0.01$, $R_{\rm m0.01}<R_{\rm c}$, theoretically, the pulsation can be formed \citep[][]{Illarionov1975}.

{\bf XSS J12270-4859 and IGR J17379-3747:} the millisecond X-ray pulsation of both XSS J12270-4859 and 
IGR J17379-3747 are discovered at the X-ray luminosity of $L_{\rm 0.5-10keV}\sim 5.0\times 10^{33} \ \rm erg \ s^{-1}$. 
The spin frequency of XSS J12270-4859 and IGR J17379-3747 are $\nu_{\rm NS}=593$ Hz and 
$\nu_{\rm NS}=468$ Hz respectively. As we can see from Table \ref{t:mdot}, at the X-ray luminosity of 
$L_{\rm 0.5-10keV}=5.0\times 10^{33} \ \rm erg \ s^{-1}$, the mass accretion rate $\dot M_{0}$ calculated with the 
formula of $\dot M_{0}=L_{\rm 0.5-10keV}{R_{*}\over {GM}}$ is $3.36\times 10^{13}\ \rm g \ s^{-1}$. 
At this X-ray luminosity, the mass accretion rate $\dot M_{0.1}$ calculated with our model of ADAF accretion 
for $f_{\rm th}=0.1$ is $7.75\times 10^{14}\ \rm g \ s^{-1}$, and the mass accretion rate $\dot M_{0.01}$ 
calculated with our model of ADAF accretion for $f_{\rm th}=0.01$ is $6.15\times 10^{15}\ \rm g \ s^{-1}$. 
%-------
Also, if we assume $M=1.4M_{\odot}$, $R_{*}=12.5\ \rm km$, $B_{*}=10^8\ \rm G$ and $\eta=0.1$
for both XSS J12270-4859 and IGR J17379-3747, the results are similar to that of PSR J1023+0038.
Specifically, if mass the mass accretion rate, i.e. $\dot M_{0}$, is calculated with the formula of 
$\dot M_{0}=L_{\rm 0.5-10keV}{R_{*}\over {GM}}$, $R_{\rm m0}>R_{\rm c}$, and if the mass accretion rate, 
i.e. $\dot M_{\rm 0.1}$, is calculated with our model of ADAF accretion for $f_{\rm th}=0.1$, $R_{\rm m0.1}>R_{\rm c}$. 
In these two cases, theoretically, the pulsation cannot be formed. If the mass accretion rate, i.e. $\dot M_{0.01}$, 
is calculated with our model of ADAF accretion for $f_{\rm th}=0.01$, $R_{\rm m0.01}<R_{\rm c}$.
In this case, theoretically, the pulsation can be formed \citep[][]{Illarionov1975}.
One can refer to Table \ref{t:radius} for the detailed numerical results of 
$R_{\rm c}$, $R_{\rm m0}$, $R_{\rm m0.1}$, and $R_{\rm m0.01}$ for XSS J12270-4859 and IGR J17379-3747 respectively.
%--------

Here, we would like to remind that we take a fixed value of the NS spin frequency, i.e., $\nu_{\rm NS}=500$ Hz,
for plotting $L_{\rm 0.5-10keV}$ as a function of $\dot m$ as in Fig. \ref{f:mdot}, and 
the corresponding calculations for $\dot M_{\rm 0.1}$ (or $\dot m_{\rm 0.1}$) and $\dot M_{\rm 0.01}$ 
(or $\dot m_{\rm 0.01}$) in Table \ref{t:mdot}, which is a little different from the observed value
of $\nu_{\rm NS}=592$ Hz for PSR J1023+0038, $\nu_{\rm NS}=593$ Hz for XSS J12270-4859, and 
$\nu_{\rm NS}=468$ Hz for IGR J17379-3747. However, we would like to remind again that it has been proven 
that the effect of the NS spin frequency, e.g., for taking $\nu_{\rm NS}=0, 200, 500$ and $700$ Hz, on the 
structure and the emergent spectra the ADAF around a NS is very little, and nearly can be neglected \citep[][]{Qiao2020a}.
So fixing the spin frequency at $500$ Hz is a good approximation for comparing with
the observational data of PSR J1023+0038, XSS J12270-4859, and IGR J17379-3747 respectively. 
Further, in our model of ADAF accretion, we do not consider the effect of the large-scale magnetic field, 
which will be discussed in Section \ref{magnetic}. 

Finally, we would like to mention that several other scenarios have been discussed for explaining the 
formation of the millisecond X-ray pulsation at the X-ray luminosity of a few times of 
$10^{33}\ \rm erg\ s^{-1}$ \citep[e.g.][]{Archibald2015,Papitto2015,Patruno2016,Bult2019}, some of which are 
summarized as follows.
(1) The coupling between the magnetic field lines and the highly conducing accretion flow maybe is weak, 
which can lead to diffusion of the gas in the accretion flow inward via Rayleigh-Taylor instability 
\citep[][]{Kulkarni2008}, or a large-scale compression of the magnetic field 
\citep[][]{Romanova2005,Ustyugova2006,Zanni2013}. In these cases, even though the magnetospheric radius is greater 
than the corotation radius, it is possible that a fraction of the gas in the accretion flow to overcome the 
centrifugal barrier of the magnetic field to be accreted onto the surface of the NS forming the X-ray pulsation.  
(2) The interaction between the magnetosphere and the accretion disc is complex, and the `propeller' effect can 
reject the infalling matter in the accretion disc only if the magnetic field at the magnetospheric radius rotates 
significantly faster than the accretion disc \citep[][]{Spruit1993}.
If this is not the case, instead, the magnetospheric radius of the accretion disc could be trapped near the
corotation radius \citep[][]{Siuniaev1977,DAngelo2010,DAngelo2012}, which has been used to explain the   
observed 1 Hz modulation in AMXP SAX J1808.4-3658 and NGC 6440 X-2 \citep[][]{Patruno2009,Patruno2013}.

\begin{table*}
\caption{The corotation radius $R_{\rm c}$ calculated with equation (\ref{e:rc}), as well as  
the magnetospheric radius $R_{\rm m0}$, $R_{\rm m0.1}$ and $R_{\rm m0.01}$ calculated with equation (\ref{e:rm}),
for PSR J1023+0038, XSS J12270-4859 and IGR J17379-3747 respectively.
Here $R_{\rm m0}$, $R_{\rm m0.1}$ and $R_{\rm m0.01}$ are calculated respectively by taking the mass accretion 
rate as $\dot M_{0}$, $\dot M_{0.1}$, and $\dot M_{0.01}$ as listed in Table \ref{t:mdot}, i.e., 
$\dot M_{0}$ is obtained with the formula of $\dot M_{0}=L_{\rm 0.5-10keV}{R_{*}\over {GM}}$, 
$\dot M_{0.1}$ are obtained from our model results of NSs with an ADAF accretion for $f_{\rm th}=0.1$,
and  $\dot M_{0.01}$ are obtained from our model results of NSs with an ADAF accretion for $f_{\rm th}=0.01$. 
In all the calculations, we take $M=1.4M_{\odot}$, $R_{*}=12.5$ km, $B_{*}=10^8$ G, and $\eta=0.1$. }
\centering
\begin{tabular}{ccccccc}
\hline
PSR J1023+0038\ \ ($\nu_{\rm NS}=592$ Hz)  \\
$L_{\rm 0.5-10keV}\ (\rm \ erg \ s^{-1})$ & $R_{\rm c}\ \rm (km)$   & $R_{\rm m0}\ \rm (km)$     & 
$R_{\rm m0.1}\ \rm (km)$   &  $R_{\rm m0.01}\ \rm (km)$     \\ 
$3.0\times 10^{33}$   &  $23.8$  &  $66.2$ & $34.2$  &  $23.1$  \\
\\
XSS J12270-4859\ \ ($\nu_{\rm NS}=593$ Hz) \\
$L_{\rm 0.5-10keV}\ (\rm \ erg \ s^{-1})$ & $R_{\rm c}\ \rm (km)$   & $R_{\rm m0}\ \rm (km)$     & 
$R_{\rm m0.1}\ \rm (km)$   &  $R_{\rm m0.01}\ \rm (km)$     \\ 
$5.0\times 10^{33}$   &  $23.8$  &  $59.8$ & $31.9$  &  $21.1$  \\
\\
IGR J17379-3747\ \ ($\nu_{\rm NS}=468$ Hz) \\
$L_{\rm 0.5-10keV}\ (\rm \ erg \ s^{-1})$ & $R_{\rm c}\ \rm (km)$   & $R_{\rm m0}\ \rm (km)$     & 
$R_{\rm m0.1}\ \rm (km)$   &  $R_{\rm m0.01}\ \rm (km)$     \\ 
$5.0\times 10^{33}$   &  $27.8$  &  $62.7$ & $33.5$  &  $22.1$  \\
\hline
\end{tabular}
\\
\label{t:radius}
\end{table*}

\section{Discussions}
\subsection{The effect of the large-scale magnetic field of $\sim 10^8\ \rm G$ on the 
radiative efficiency of NSs with an ADAF accretion}\label{magnetic}
In this paper, we investigate the radiative efficiency of weakly magnetized NSs with an ADAF accretion 
for taking two typical values of $f_{\rm th}=0.1$ and $f_{\rm th}=0.01$ as suggested 
in \citet[][]{Qiao2020a} and \citet[][]{Qiao2020b}. Then, we show that NSs with an ADAF 
accretion is radiatively inefficient, with which we further explain the observed millisecond 
X-ray pulsations for PSR J1023+0038, XSS J12270-4859 and IGR J17379-3747 at the X-ray luminosity of a 
few times of $10^{33}\ \rm erg\ s^{-1}$. However, we should note that, in our model of NSs with an 
ADAF accretion, we do not consider the effect of the large-scale magnetic field on the emission 
of the ADAF, which probably will affect the radiative efficiency of the NSs with an ADAF accretion. 

In general, accreting millisecond X-ray pulsars (AMXPs) are believed to have a relatively weaker
magnetic field of $\sim 10^8$ G \citep[e.g.][for review]{Wijnands1998,Casella2008,Patruno2012}, which is 
different from the standard X-ray pulsars often with a stronger magnetic field of 
$\sim 10^{12}$ G \citep[e.g.][for review]{Coburn2002,Pottschmidt2005,Caballero2012,Revnivtsev2015}. 
{Due to the relatively weaker magnetic field in AMXPs}, it is often suggested that the magnetic 
field in AMXPs does not significantly affect the X-ray spectra \citep[e.g.][]{Poutanen2003}, which seems 
to be supported by some observations by comparing the X-ray spectra between the non-pulsating NS-LMXBs 
and the AMXPs. {In general, it is found that there is no systematic difference of the X-ray spectra    
between the non-pulsating NS-LMXBs and the AMXPs 
in the range of $L_{\rm 0.5-10\rm keV} \sim 10^{34}-10^{36} \rm \ erg \ s^{-1}$.}
%------------
For example, in \citet[][]{Wijnands2015}, the author compiled a sample composed of eleven non-pulsating
NS-LMXBs, finding that systematically there is an anticorrelation  
between the X-ray photon index $\Gamma$ (obtained by fitting the X-ray spectra between 0.5 and 10 keV 
with a single power law) and the X-ray luminosity $L_{\rm 0.5-10keV}$ in the range of 
$L_{\rm 0.5-10\rm keV} \sim 10^{34}-10^{36} \rm \ erg \ s^{-1}$. Further, the authors added three 
AMXPs, i.e., NGC 6440 X-2, IGR J00291+5934, and IGR J18245-2452, with well measured $\Gamma$ and 
$L_{\rm 0.5-10keV}$ to compare with the non-pulsating NS sample, showing that at a fixed X-ray 
luminosity, the X-ray spectra of the AMXPs appear to be slightly harder than that of the non-pulsating 
NS-LMXBs.
%------------
More accurately, the authors did 2D KS test to study whether the AMXP data are consistent with
the non-pulsating data. It is found that a 90 per cent confidence interval for the probability
of $1.2\times 10^{-6}-3.5\times 10^{-4}$ that the AMXP data and the non-pulsating data have the same 
distribution. However, given the fact that only three AMXPs are included in this study, actually, the authors 
also reminded that they cannot draw strong conclusions whether the presence of the magnetic field in AMXPs can 
alter the X-ray spectra \citep[][]{Wijnands2015}.    
%------------
{In a further study of \citet[][]{Parikh2017}, the authors combined} the data 
in \citet[][]{Wijnands2015} and some additional new data in the range of 
$L_{\rm 0.5-10\rm keV} \sim 10^{34}-10^{36} \rm \ erg \ s^{-1}$
for the anticorrelation between the X-ray photon index $\Gamma$ and the X-ray luminosity $L_{\rm 0.5-10\rm keV}$, 
the authors showed that they did not find that the X-ray spectra of AMXPs are systematically harder than that of 
the non-pulsating sources as tested in \citet[][]{Wijnands2015}, suggesting that the hardness of the X-ray spectra 
does not have strict connection with the presence of the dynamic effect of the magnetic field.

{As for $L_{\rm 0.5-10\rm keV} \lesssim 10^{34} \rm \ erg \ s^{-1}$ (generally defined as 
the quiescent state), the X-ray spectra of non-pulsating 
NS-LMXBs are very complex and diverse, which can be (1) completely dominated by a thermal soft X-ray 
component, (2) completely dominated by a power-law component, or (3) described by the two-component 
model, i.e. a thermal soft X-ray component plus a power-law component \citep[e.g.][for discussions]{Wijnands2015}.
For example, the X-ray spectra of the non-pulsating NS-LMXB Cen X-4 at the X-ray luminosity of 
$L_{\rm 0.5-10\rm keV} \sim 10^{33} \rm \ erg \ s^{-1}$ can be well fitted by the two-component model, 
i.e. a thermal soft X-ray component plus a power-law component, revealing a harder X-ray photon index of
$\Gamma \sim 1-1.5$ \citep[][]{Chakrabarty2014,DAngelo2015}, 
while the X-ray spectra of several non-pulsating NS-LMXBs are well fitted by a single power law with a 
softer X-ray photon index of $\Gamma \sim 3-5$ at the X-ray luminosity of  
$L_{\rm 0.5-10\rm keV}\sim$ a few times of $10^{33} \rm \ erg \ s^{-1}$ \citep[][]{Sonbas2018}.
%------------
For the three sources, i.e., PSR J1023+0038, XSS J12270-4859 and IGR J17379-3747 with the 
millisecond X-ray pulsations observed at the X-ray luminosity of  
$L_{\rm 0.5-10\rm keV}\sim$ a few times of $10^{33} \rm \ erg \ s^{-1}$, it is found that the X-ray spectra 
can be well fitted by a single power law with the photon index $\Gamma\sim 1.7$ for PSR J1023+0038 
\citep[][]{Archibald2015}, with $\Gamma\sim 1.6$ for XSS J12270-4859 \citep[][]{Saitou2009}, 
and can be well fitted by two thermal components, i.e., a thermal component of $\sim 0.35$ keV 
plus a thermal component of $\sim 0.12$ keV for IGR J17379-3747, indicating a very soft X-ray 
spectrum (the Group 3 data) \citep[][]{Bult2019}.}

%------------
In summary, as discussed above we think that the effect of the magnetic field of $\sim 10^8$ G in AMXPs on 
the emission of NSs with an ADAF accretion in the range of 
$L_{\rm 0.5-10\rm keV} \sim 10^{34}-10^{36} \rm \ erg \ s^{-1}$ is very little, 
consequently the effect of the magnetic field on the radiative efficiency of NSs with an ADAF accretion is 
very little. As for at the X-ray luminosity of $L_{\rm 0.5-10\rm keV}\sim$ a few times of $10^{33} \rm \ erg \ s^{-1}$,
we think it is not very easy to say whether there is significant effect of the magnetic field of 
$\sim 10^8$ G on the emission of NSs with an ADAF accretion. Here, at least for PSR J1023+0038 and 
XSS J12270-4859, if the X-ray spectra (well described by a single power law) can be explained by our model ADAF 
accretion, it requires a very small value of $f_{\rm th}$ (for decreasing the contribution of the thermal soft 
X-ray component), i.e., $f_{\rm th}$ approaching to 
zero (even smaller than 0.01 as taken in the present paper) [see Fig. 7 in \citet[][]{Qiao2020a} for details]. 
So we think that our explanations for the observed millisecond X-ray pulsations at the X-ray luminosity of 
a few times of $10^{33}\ \rm erg\ s^{-1}$ with our model of ADAF accretion by taking a small value 
of $f_{\rm th}$, i.e. $f_{\rm th}=0.01$ is a good approximation.
Here, we would like to address that due to the existence of the magnetic field of $\sim 10^8$ G, 
the boundary condition in the region between the surface of the NS and the ADAF in AMXPs should be different from  
that of in non-pulsating NSs, which however has been incorporated into the
effect of the parameter $f_{\rm th}$ if we only focus on this question from the viewpoint of emission.
%------------
Finally, we also would like to address that a detailed study of the effects of the large-scale magnetic 
field of $\sim 10^8$ G on the dynamics and the emission of NSs with an ADAF accretion is still very necessary 
for the consistency between the model and the observations for AMXPs in the future, although the effects of 
the magnetic field at the strength of $\sim 10^8$ G on the radiative efficiency of NSs with an ADAF accretion 
maybe are not very obvious. 

\subsection{Further observational test for the radiative efficiency of NSs with an ADAF accretion in the future}
In our model of NSs with an ADAF accretion, there is a very important parameter, $f_{\rm th}$, which controls 
the feedback between the surface of the NS and the ADAF. The value of $f_{\rm th}$ can affect the radiative 
efficiency of NSs with an ADAF accretion. As has been shown in \citet[][]{Qiao2020a} and \citet[][]{Qiao2020b}, 
the value of $f_{\rm th}$ has been constrained to be less than 0.1, and it seems that a smaller value 
of $f_{\rm th}$, i.e., $f_{\rm th}\sim 0.01$ is more preferred.
It is possible that the remaining fraction, i.e., 1-$f_{\rm th}$, of the ADAF energy transferred onto the 
surface of the NS could be partially converted to the rotational energy of the NS, and could be partially 
absorbed by the NS and stored as the internal energy at the crust of the NS.
The accreted matter in the form of the ADAF (with relatively higher temperature and lower density) and 
the carried energy itself may produce some additionally observational features at the surface of the NS, which 
currently however has not been well investigated, depending on the resulted changes of the 
temperature and the density of the matter in the very thin layer at the surface of the 
NS \citep[e.g.][for the related discussions]{Galloway2021}.
The study of the further effects of the accreted matter in the form of the ADAF at the surface of 
the NS exceeds the research scope in the present paper, and definitely will be carried out in the future.

In \citet[][]{Qiao2020a} and \citet[][]{Qiao2020b}, the constraint to the value of $f_{\rm th}$ is based on some 
statistically observed correlations in non-pulsating NSs, such as the fractional contribution of the power-law 
component $\eta$ as a function of $L_{\rm 0.5-10keV}$, as well as the X-ray photon index $\Gamma$ as a function of 
$L_{\rm 0.5-10keV}$. In order more precisely to constrain the value of $f_{\rm th}$, we expect that the detailed 
X-ray spectral fittings will be done for some typically single source in the future, such as the study for
Cen X-4 \citep[e.g.][]{Chakrabarty2014, DAngelo2015}. 

As discussed in Section \ref{s:pulsation}, if our model of ADAF accretion can be applied to explain 
the observed millisecond X-ray pulsation at the X-ray luminosity of a few times of $10^{33}\ \rm erg\ s^{-1}$ 
for PSR J1023+0038, XSS J12270-4859 and IGR J17379-3747, a small value of $f_{\rm th}$, e.g., 
$f_{\rm th}=0.01$ is required. {Based on some related results from the model of ADAF accretion 
for taking $f_{\rm th}=0.01$, we can further estimate the change rate of the NS spin frequency 
$\dot \nu_{\rm NS}$.}
%------------
{If we assume that the change of the NS spin is due to the accretion, according to the 
conservation of angular momentum, we have}
\begin{eqnarray}\label{e:angularm} 
I\dot \Omega_{\rm NS}=\dot M(\Omega_{*}-\Omega_{\rm NS})R_{*}^2,
\end{eqnarray}
where $I$ is the moment of inertia of the NS, $\Omega_{*}$ is the rotational angular velocity of the ADAF at $R_{*}$
with $\Omega_{*}=2\pi\nu_{*}$ (with $\nu_{*}$ being the angular frequency at $R_{*}$), and $\Omega_{\rm NS}$ is the 
rotational angular velocity of the NS with $\Omega_{\rm NS}=2\pi\nu_{\rm NS}$. 
Rearranging equation (\ref{e:angularm}), we can express the change rate of the NS spin frequency as follows, 
\begin{eqnarray}\label{e:frequencym} 
\dot \nu_{\rm NS}={\dot M(\nu_{*}-\nu_{\rm NS})R_{*}^2}/I.
\end{eqnarray}
Given the value of $\dot M$, $\nu_{*}$, $\nu_{\rm NS}$ and $I$, we can calculate the change rate of the NS spin 
frequency $\dot \nu_{\rm NS}$.
For example, for PSR J1023+0038 the millisecond X-ray pulsation is observed at the X-ray luminosity of 
$L_{\rm 0.5-10keV}\sim 3\times 10^{33}\ \rm erg\ s^{-1}$, the corresponding $\dot M$ is 
$3.88\times 10^{15}\ \rm g \ s^{-1}$ based on our model of ADAF accretion for $f_{\rm th}=0.01$. 
With $\dot M=3.88\times 10^{15}\ \rm g \ s^{-1}$, we recalculate the structure of the ADAF for $\nu_{*}$. 
The value of $\nu_{*}$ is 253 Hz. The moment of inertia $I$ is $\sim 1.75\times 10^{45}\ \rm g\ cm^2$ for 
taking the typical value of $M=1.4M_{\rm \odot}$ and  $R_{*}=12.5\ \rm km$ respectively.
The spin frequency $\nu_{\rm NS}$ of PSR J1023+0038 is $592$ Hz.
Substituting the value of $\dot M$, $\nu_{*}$, $\nu_{\rm NS}$ and $I$ into equation (\ref{e:frequencym}),
we get $\dot \nu_{\rm NS}\sim -1.2\times 10^{-15}\ \rm Hz\ s^{-1}$,  
which is close to ($\sim 2.5$ times less than) the observed value of 
$\dot \nu_{\rm NS}\sim -3.04 \times 10^{-15}\ \rm Hz\ s^{-1}$ for PSR J1023+0038 at the LMXB 
state \citep[][]{Jaodand2016}. 
Here, we should note that in this case, the value of $\nu_{*}$ from our model of ADAF accretion is 
less than $\nu_{\rm NS}$, which means that a negative torque will be exerted on the NS, consequently making 
the rotational energy of the NS transferred onto the ADAF and the NS to be spin-down, rather than the ADAF 
energy transferred onto the NS and the NS to be spin-up. Further, since a variable flat-spectrum of radio 
emission is revealed as PSR J1023+0038 in the LMXB state, it means that the outflow is existed, which 
physically can further make the NS to be spin-down to match the observed 
$\dot \nu_{\rm NS}$ \citep[][]{Deller2015}. 

A similar calculation for $\dot \nu_{\rm NS}$ is done for XSS J12270-4859 and IGR J17379-3747
with the millisecond X-ray pulsations observed at the X-ray luminosity of 
$\sim 5\times 10^{33}\ \rm erg\ s^{-1}$. At this X-ray luminosity, the mass accretion rate $\dot M$ is 
$6.15\times 10^{15}\ \rm g \ s^{-1}$ based on our model of ADAF accretion for $f_{\rm th}=0.01$.
With $\dot M=6.15\times 10^{15}\ \rm g \ s^{-1}$, we recalculate the structure of the ADAF for $\nu_{*}$.
The value of $\nu_{*}$ is 278 Hz. The spin frequency $\nu_{\rm NS}$ is $593$ Hz for XSS J12270-4859, and 
is $468$ Hz for IGR J17379-3747. The moment of inertia $I$ is $\sim 1.75\times 10^{45}\ \rm g\ cm^2$ 
for taking $M=1.4M_{\rm \odot}$ and $R_{*}=12.5\ \rm km$ respectively.
Again substituting the value of $\dot M$, $\nu_{*}$, $\nu_{\rm NS}$ and $I$ into equation (\ref{e:frequencym}),
we get $\dot \nu_{\rm NS}\sim -1.73\times 10^{-15}\ \rm Hz\ s^{-1}$ for XSS J12270-4859 and 
$\dot \nu_{\rm NS}\sim -1.0\times 10^{-15}\ \rm Hz\ s^{-1}$ for IGR J17379-3747.
It is clear that the value of $\dot \nu_{\rm NS}$ is negative (i.e., spin-down) for  
XSS J12270-4859 and IGR J17379-3747 as for PSR J1023+0038, which means that the rotational energy of the NS is 
transferred onto the ADAF. 
%-----------
If our explanation for the formation of the observed millisecond X-ray pulsations  
for XSS J12270-4859 and IGR J17379-3747 at the X-ray luminosity of $5\times 10^{33}\ \rm erg\ s^{-1}$ are 
correct, the predicted change rate of the NS spin frequency $\dot \nu_{\rm NS}$ is at the level of 
$\sim -10^{-15}\ \rm Hz\ s^{-1}$, which we expect can be tested by the observations in the future.
Further, if the change rate of the NS spin frequency $\dot \nu_{\rm NS}$ predicted by our model of ADAF 
accretion can be confirmed in the future, which actually in turn supports our idea in the present 
paper that NSs with an ADAF accretion is radiatively inefficient despite the existence of the hard surface.  
Finally, we would like to address that the estimation of $\dot \nu_{\rm NS}$ in this paper is
based on our model of ADAF accretion around a weakly magnetized NS, which will make the estimated 
value of $\dot \nu_{\rm NS}$ uncertain as applied to the AMXP cases. So the consideration of the effect of 
the magnetic field ($\sim 10^8$ G) on the value of $\dot \nu_{\rm NS}$ in AMXPs is still very necessary 
in the future, which however exceeds the scope in the present paper. 

\section{Conclusions}
Following the paper of \citet[][]{Qiao2020a} and \citet[][]{Qiao2020b} for the constraints to the value 
of $f_{\rm th}$ controlling the feedback between the surface of the NS and the ADAF, in this paper, we investigate 
the radiative efficiency of NSs with an ADAF accretion within the framework of the self-similar solution of the ADAF
by taking two typically suggested values of $f_{\rm th}$, i.e., $f_{\rm th}=0.1$ and $f_{\rm th}=0.01$ respectively.
Then, we show that the radiative efficiency of NSs with an ADAF accretion is significantly lower 
than that of $\epsilon \sim {\dot M GM\over R_{*}}/{\dot M c^2}\sim 0.2$.
Specifically, the radiative efficiency of our model of NSs with an ADAF accretion for $f_{\rm th}=0.1$ is 
roughly one order of magnitude lower than that of $\epsilon \sim {\dot M GM\over R_{*}}/{\dot M c^2}\sim 0.2$,
and the radiative efficiency of our model of NSs with an ADAF accretion for $f_{\rm th}=0.01$ is roughly two 
orders of magnitude lower than that of $\epsilon \sim 0.2$.
As a result, we propose that the lower radiative efficiency of our model of ADAF accretion probably can be applied to
explain the observed millisecond X-ray pulsation in some NS-LMXBs (such as PSR J1023+0038, XSS J12270-4859
and IGR J17379-3747) at the X-ray luminosity (between 0.5 and 10 keV) of a few times of $10^{33}\ \rm erg\ s^{-1}$, 
since at this X-ray luminosity the real $\dot M$ calculated with our model of ADAF accretion for taking 
an appropriate value of $f_{\rm th}$, such as $f_{\rm th}=0.01$, can be more than two orders of magnitude higher 
than that of calculated with the formula of $\dot M=L_{\rm X}{R_{*}\over {GM}}$ to ensure a fraction of the matter 
in the ADAF to be channelled onto the surface of the NS forming the X-ray pulsation.  

%-------------------------------------------------------------------------------------
\section*{Acknowledgments}
Erlin Qiao thanks the very useful discussions with Dr. Chichuan Jin from NAOC.
This work is supported by the National Natural Science Foundation of 
China (Grants 11773037 and 11673026), the gravitational wave pilot B (Grant No. XDB23040100), 
the Strategic Pioneer Program on Space Science, Chinese Academy of Sciences (Grant No. XDA15052100), 
and the National Program on Key Research and Development Project (Grant No. 2016YFA0400804).

\section*{Data availability}
The data underlying this article will be shared on reasonable request to the
corresponding author.

\bibliographystyle{mnras}
\bibliography{qiaoel}

%\begin{thebibliography}{99}
%\end{thebibliography}

%----------------------------------------------------
% Don't change these lines
\bsp	% typesetting comment
\label{lastpage}
\end{document}